\documentclass[12pt]{article}
\usepackage{graphicx}
\usepackage{bm}
\usepackage[version=4]{mhchem}
\usepackage[scientific-notation=true]{siunitx}
\usepackage{cite}
\usepackage{authblk}

\begin{document}

\title{A Predictive Model for Convective Flows Induced by Surface Reactivity Contrast}

\author[1]{Scott M. Davidson}
\author[2]{Rob G.H. Lammertink}
\author[1]{Ali Mani\thanks{Corresponding author: alimani@stanford.edu}}
\affil[1]{Department of Mechanical Engineering, Stanford University, Stanford, CA 94305, USA}
\affil[2]{Soft Matter, Fluidics and Interfaces, Mesa+ Institute for Nanotechnology, University of Twente, 7500 AE Enschede, The Netherlands}

\maketitle

\begin{abstract}
Concentration gradients in a fluid along a reactive surface due to contrast in surface reactivity generate convective flows. These flows result from contributions by electro and diffusio osmotic phenomena. In this study we have analyzed reactive patterns that release and consume protons, analogous to bimetallic catalytic conversion of peroxide. Here, we present a simple analytical model that accurately predicts the induced potentials and consequent velocities in such systems over a wide range of input parameters. Our model is tested against direct numerical solutions to the coupled Poisson, Nernst-Planck, and Navier-Stokes equations. Our analysis can be used to predict enhancement of mass transport and the resulting impact on overall catalytic conversion, and is also applicable to predicting the speed of catalytic nanomotors.
\end{abstract}


\section{Introduction}
The ability of catalytic micropumps to generate motion from chemical energy has opened many possibilities~\cite{TimothyR.Kline2005, Paxton2006a, Zhou2016}. By generating convection at the microscale, catalytic micropumps can enhance transport rates to surfaces where diffusion would otherwise be limiting~\cite{Squires2008}. Due to their use of local chemical energy, catalytic micropumps don't suffer from some of the issues associated with other methods of generating flow at the microscale. Pressure-driven flow is inefficient due to the rapid increase in hydraulic resistance with decreasing length scale. Meanwhile other electrokinetic methods such as conventional electroosmotic flow and induced-charge electroosmosis~\cite{Bazant2004, Squires2004} require imposition of external electric fields or, in the case of alternating-current electroosmosis, electrical connection of micro-electrodes~\cite{Ramos1998, Ramos1999, Ramos2003}.

Surfaces with nonuniform catalytic reactivity generate flow by multiple mechanisms~\cite{Anderson1989}. One mechanism, electroosmosis, refers to flows generated when reaction-induced electric fields, driving current through the fluid, act on surface electric double layers screening the surface charge. This mechanism has been studied using a silver disc plated on a gold substrate immersed in a hydrogen peroxide solution~\cite{TimothyR.Kline2005, TimothyR.Kline2006} as well as for interdigitated gold and platinum electrodes~\cite{Paxton2006}. Ibele \textit{et al.} showed that different fuels can generate flow by the same mechanism by developing a system using hydrazine fuels~\cite{Ibele2007}. Farniya and coworkers developed a technique for inferring chemical reaction rates and ion impurities in electrocatalytic micropumps by fitting simulations to detailed experiments measuring proton concentrations with fluorescence microscopy and tracking of particles with positive and negative surface charges to measure flow velocities and electric fields~\cite{Farniya2013}. In other experiments it has been shown that separating the electrodes on opposite sides of a membrane~\cite{Jun2010} and utilizing the high surface charge of doped silicon provide promising avenues for enhancing the flow rates generated by electrocatalytic micropumps~\cite{Esplandiu2015}. Recent simulations by Esplandiu \textit{et al.} analyzed important parameters in self-electroosmotic pumps and suggested that the induced electric fields are dependent on ionic strength, not total conductivity~\cite{Esplandiu2016}. However their model assumed that only protons contribute to the net current and ignored bulk equilibrium reactions.

A second mechanism, diffusioosmosis, refers to flows generated by the presence of species concentration gradients tangent to the surface. When the concentration gradients are caused by nonuniform reactivity of the surface itself the flows they generate are called self-diffusioosmotic. Such gradients may generate flows in multiple ways depending on whether the solution is an electrolyte~\cite{Anderson1982} or not~\cite{Prieve1984}. In electrolyte solutions, gradients of species concentrations generate diffusioosmotic flows due to the dependence of the pressure difference across the Debye screening layer on solution conductivity. The surface pressure is proportional to the local conductivity, so concentration gradients along the surface generate pressure gradients and hence flow. Also, for electrolytes where the positive and negative species have different diffusivities, concentration gradients cause internal electric fields which act on the Debye layer to generate flow. Diffusioosmotic flows have been used to develop microfluidic pumps~\cite{Zhou2016}. Hong \textit{et al.} utilized the photocatalytic sensitivity of titanium-dioxide to generate a diffusioosmotic micropump controllable by light~\cite{Hong2010}. Niu \textit{et al.} showed that even in micromolar salt concentrations, an ion-exchange resin can be used to create a self-diffusioosmotic micropump~\cite{Niu2017}. 


Although we are primarily concerned with electrocatalytic micropumps, much research into electrocatalytic conversion of chemical to mechanical energy has focused on reaction induced propulsion of nano-swimmers i.e. self-electrophoresis and self-diffusiophoresis. The same mechanisms which cause pumping on a fixed surface lead to swimming of a particle with varying surface reactivity. Such self-phoretic mechanisms for motion were originally investigated in the context of cell motion~\cite{Mitchell1972, Lammert1996}. Early experiments demonstrating reaction-induced propulsion in non-living systems utilized bimetallic particles in peroxide solutions~\cite{Fournier-Bidoz2005, Paxton2005, Wheat2010}. Various proposed mechanisms explaining their motion were proposed, and of these self-electrophoresis was found to be dominant~\cite{Paxton2006, YangWang2006}. Velocities on the order 10~$\mu$m/s were reported. Nanomachines such as these have created much excitement due to their potential to revolutionize fields such as drug delivery and biological sensing~\cite{Wang2009, Wu2010, Ebbens2010, Sengupta2012}.


A number of theoretical studies have been performed to better understand self-catalytic swimming.  Golestanian showed how asymmetric reaction induced concentration gradients result in the propulsion of particles by self-diffusiophoretic mechanisms \cite{Golestanian2005}. Moran and Posner investigated the self-electrophoretic mechanism by performing direct numerical simulations of the bulk transport equations coupled initially to fixed surface fluxes~\cite{Moran2010} which they refined by considering surface reaction kinetics~\cite{MORAN2011}. They also analyzed the reduction in swimming speed with background electrolyte concentration~\cite{Moran2014}. However, their model predicted a quadratic relationship between swimming speed and peroxide concentration unlike the linear dependence observed in experiments. Yariv studied the problem using matched asymptotic expansions in the limit of thin electrical double layers and long-slender rods, predicting propulsion in experimentally observed direction~\cite{Yariv2011}. Sabass and Seifert found that in the presence of background salt, the decrease in pH with increased peroxide concentration can alleviate diffusive limitations on proton transport~\cite{Sabass2012}. Self-consistent nonlocal feedback theory was developed to further improve understanding of electrocatalytic motors~\cite{Nourhani2015}. Further theoretical studies have found that the optimal location for surface reactivity is concentrated at the poles of the swimmer~\cite{Kreissl2016}, and that bulk peroxide decomposition can significantly affect swimming speeds~\cite{Brown2017}.

The aforementioned studies contribute identification of mechanisms leading to electrostatic potential gradients and fluid flow in systems involving surface reactivity contrasts. However, a quantitative understanding that would directly connect the outcome of induced potentials and flow fields to the key input parameters is yet to be developed. Ideally, it would be desirable to have an algebraic model predicting this outcome. As we shall see, flow fields in these systems can have a nonmonotonic dependence on input and therefore it is crucial to have a model that goes beyond simple scaling relations in large and small parameter limits. Our study presents such models for a canonical problem.

Here, we investigate the relation between surface reactivity patterns and the fluid flow they generate. For this, we analyzed the scaling relation of the induced potential (gradient) and velocity with surface reactivity pattern. The system corresponds to a pattern of catalytic sites with distinct reaction rate constants (expressed by their corresponding second Damk{\"o}hler numbers). We provide physical insight into how the different components of the induced potential lead to a balance of fluxes in the system. We also derive a simple model capturing the induced potential and flow velocity and compare this with direct numerical simulations of the governing transport equations.


%
%

\section{Model Problem}
Many combinations of patterned metals and electrolytes lead to reactions which generate the asymmetric ion concentrations necessary to induce this type of flow~\cite{YangWang2006}. In this work, we focus on  electrocatalytic hydrogen peroxide oxidation at Pt next to its reduction on Au surfaces. The oxidation generates electrons and protons, while the reduction consumes protons and electrons. This asymmetric surface reaction pair generates and depletes protons near the surface resulting in a reaction induced charge distribution and electric field. The forces on the fluid caused by the induced electric field acting on the induced charge distribution gives rise to an electroosmotic flow which is driven by the electrocatalytic conversion.
\begin{figure}[htb!]
	\centering
    \includegraphics[width=.5\columnwidth]{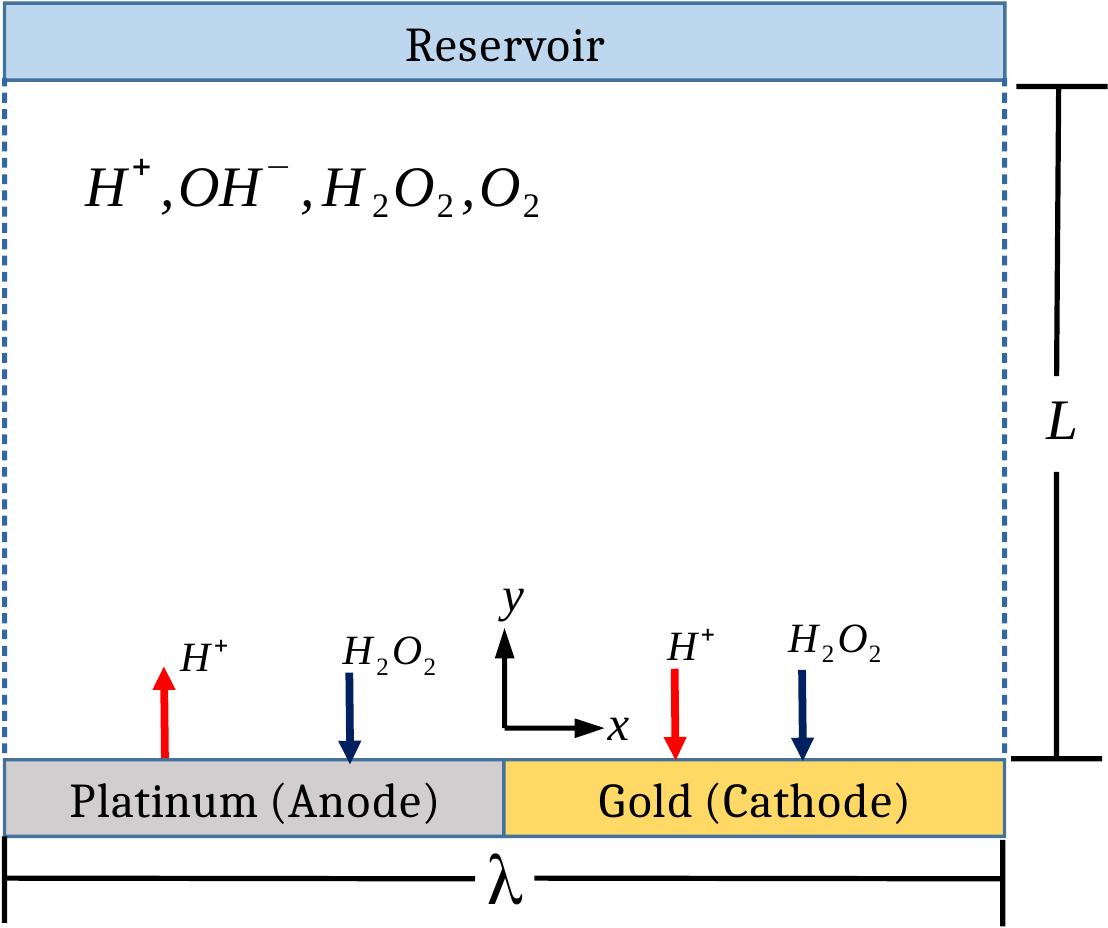}
    \caption{(Color online) Schematic showing the model problem. The electrocatalytic surface reaction pair consists of Gold (Au) and Platinum (Pt) at a periodic width of $\lambda$ in contact with the electrolyte containing protons, hydroxide ions, hydrogen peroxide and oxygen.}
	\label{fig:schematic}
\end{figure}

Specifically, we model a two-dimensional domain of width $\lambda$ and height $L$ consisting of an electrolyte between a reactive surface and a reservoir as shown in Figure (\ref{fig:schematic}). The domain is periodic in the tangential $x$ direction, and thus $\lambda$ represents the period of a repeating patterned surface of which we simulate one period. This surface is modeled as completely flat. The background electrolyte contains four species: protons \ce{H^+}, hydroxide ions \ce{OH-}, hydrogen peroxide \ce{H2O2}, and molecular oxygen \ce{O2}. Peroxide oxidation at the platinum anode is given by
\begin{align}
\label{eqn:pt_red}
\ce{H2O2 -> 2H+ + O2 + 2e-}
\end{align}
and reduction at the gold cathode is given by
\begin{align}
\ce{H2O2 + 2H+ + 2e- -> 2H2O}
\end{align}
In this work, we only explicitly simulate the concentrations of the two ionic species, protons and hydroxide ions. We assume that the concentration of peroxide is constant over relevant timescales so its transport can be ignored. This assumption is justified with scaling arguments in Section (\ref{sec:two_species}). Transport of oxygen can be ignored because in the dilute limit oxygen only affects the system through the anode reactions. However, we assume that the anode reaction only proceeds in the forward direction in which oxygen does not participate~\cite{MORAN2011}.

\subsection{Governing Equations}

This system is governed by the coupled Poisson-Nernst-Planck (PNP) equations for ion transport and Navier-Stokes (NS) equations of fluid motion along with appropriate surface and bulk reactions. The PNP equations relate species concentration change to advection, diffusion, reaction, and electromigration fluxes. Transport of ionic species is coupled through the Poisson equation for the electric potential. In dimensionless form they are given by
\begin{align}
\frac{\partial c_i}{\partial t} + \nabla \cdot \left( \bm{u}c_i \right) &= D_i \nabla \cdot \left( \nabla c_i + z_i c_i \nabla \phi \right) + R_i \\
-2\epsilon^2 \nabla^2 \phi &= \sum_{i} z_i c_i
\end{align}
where $c_i = \widetilde{c_i}/c_0$ is the dimensionless concentration of species $i$ which has valence $z_i$ and diffusivity $D_i = \widetilde{D_i}/D_0$, where the $\,\widetilde{}\,$ represent dimensional quantities when there is a dimensionless counterpart. $c_0$ and $D_0$ are arbitrary reference values for concentration and diffusivity. In this study we reference diffusivities to the diffusivity of hydroxide ions $D_0=\num{5.273e-9}\, \mathrm{m^2/s}$~\cite{MORAN2011} and concentrations to the bulk concentrations of \ce{H+} and \ce{OH-} in pure water, $c_0=\sqrt{K_w}=\num{1e-7} \mathrm{M}$ where $K_w$ is the water equilibrium constant. $\phi = \widetilde{\phi}/V_T$ is the electrostatic potential normalized by the thermal voltage $V_T=RT/F\approx25\, \mathrm{mV}$ where $R$ is the ideal gas constant, $T$ the temperature, and $F$ the Faraday constant. $\bm{u}=\bm{\widetilde{u}}/(D_0/\lambda)$ is the fluid velocity normalized by the diffusion velocity. $t=t/t_0$ where $t_0=\lambda^2/D_0$ is the diffusion time. The coefficient $\epsilon$ is the dimensionless Debye length $\epsilon = \lambda_D/\lambda$ where for our case of a symmetric binary electrolyte $\lambda_D = \sqrt{\varepsilon V_T/2Fc_0}$ where $\varepsilon$ is the permittivity of water. $R_i=\widetilde{R_i}/(c_0/t_0)$ is a bulk reaction term. For hydrogen peroxide and molecular oxygen, $R_i = 0$. For protons and hydroxide ions, $R_i = Da_b\left(K_w/c^2_{0} - c_+c_-\right)$ which enforces water self-ionization. $Da_b=k_b\lambda^2c_0/D_0$ is a dimensionless bulk Damk{\"o}hler number where $k_b$ is the bulk water recombination reaction rate constant and $c_+$ and $c_-$ are proton and hydroxide ion concentrations respectively. With this nondimensionalization, all lengths are normalized by $\lambda$ and thus the dimensionless width is $1$. We chose an aspect ratio of $L/\lambda=4$, which is large enough that the reservoir boundary conditions do not affect the simulated flow near the electrodes.

The Navier-Stokes equations for momentum conservation along with continuity which enforces the incompressibility of water are
\begin{align}
\label{eqn:NS}
\frac{1}{Sc}\frac{\partial \bm{u}}{\partial t} &= -\nabla p + \nabla^2 \bm{u} - \frac{\kappa}{2\epsilon^2}\rho_e \nabla \phi\\
\nabla \cdot \bm{u} &= 0
\end{align}
Here we have ignored the nonlinear inertial term due to the small Reynolds number of the problem and included the electric body force term. $p=\widetilde{p}/(\mu D_0/\lambda^2)$ is the pressure, and $\rho_e=\sum_{i}z_i c_i/c_0$ the charge density. The Schmidt number $Sc=\mu/\rho D_0$ where $\mu$ is the dynamic viscosity and $\rho$ the density of water, and the electrohydrodynamic coupling constant $\kappa=\varepsilon V_T^2/\mu D_0$. Note that we do not include any chemiosmotic forces (i.e.\ forces due to nonionic solute-wall interactions) because gradients in nonionic solutes (O$_2$ and H$_2$O$_2$) are expected to be very small. However, the diffusioosmotic forces due to internal electric fields caused by salt concentration gradients with ions of different diffusivities are naturally included by this formulation.

\subsection{Boundary Conditions}

The boundary condition for the NS equations at both the electrodes $\left(y=0\right)$ and reservoir $\left(y=4\right)$ is given by
\begin{align}
\bm{u} = \bm{0}
\end{align}
due to the no-slip and no-penetration conditions at the solid surface along with our assumption of a stationary reservoir.

The potential boundary condition at the electrodes is given by
\begin{align}
\phi = 0
\end{align}
thus representing the constant electric potential of a conductive metal surface which we have arbitrarily set to zero as a reference. Due to the thinness of the Stern layer to the Debye length for the dilute conditions studied, we omit a Stern layer correction~\cite{MORAN2011, Esplandiu2016}. The potential boundary condition at the reservoir is
\begin{align}
\frac{\partial \phi}{\partial y} = 0
\end{align}
or zero normal electric field. As with the reservoir boundary condition for the NS equations, this condition is valid for sufficiently large $L/\lambda$ such that, in the absence of external electric fields, electric fields induced by the reactions decay before the reservoir.

For species transport, anions do not participate in either of the surface reactions, so a no-flux boundary condition is enforced on the entire surface. This reduces to
\begin{align}
\frac{\partial c_{-}}{\partial y} - c_{-}\frac{\partial \phi}{\partial y} = 0
\end{align}

For cations, different boundary conditions are used for the platinum and gold portions of the surface. Both are derived by matching the combined electromigration and diffusion fluxes with the surface reaction flux. The platinum ($-1/2 < x < 0$) and gold ($0 < x < 1/2$) boundary conditions respectively are
\begin{align}
-D_{+}\left(\frac{\partial c_{+}}{\partial y} + c_{+}\frac{\partial \phi}{\partial y}\right)&=D_{+} Da_{a}\\
-D_{+}\left(\frac{\partial c_{+}}{\partial y} + c_{+}\frac{\partial \phi}{\partial y}\right)&=-D_{+} Da_{c} c^2_{+}
\end{align}
where $Da_{a}=k_{0,\,anode}\lambda\widetilde{c}_{H_2O_2}/\widetilde{D_{+}}c_0$ is the anode Damk{\"o}hler number and $Da_{c}=k_{r,\,cathode}\lambda\widetilde{c}_{H_2O_2}c_0/\widetilde{D_{+}}$ is the cathode Damk{\"o}hler number defining the dimensionless anode and cathode reaction rates. Both of these boundary conditions assume that the reactions proceed only in the forward direction and that the Stern layer potential drops are sufficiently small that reaction rate coefficients can be treated as constants.

The reservoir boundary condition for both species is simply a Dirichlet condition at the reservoir concentration
\begin{align}
c_+=c_-= 1
\end{align}

The left and right boundary are treated as periodic boundaries.

\subsection{Validity Regime of Two-Species Approximation} \label{sec:two_species}
Our assumption that we can treat the concentration of \ce{H2O2} as a constant is valid so long as the timescale associated with changes in the concentration of \ce{H2O2} is much longer than dynamically relevant timescales of the system. To estimate this timescale, we start with the balance of reactive and diffusive fluxes at the anode surface. The scaling of these fluxes can be written in dimensionless form as
\begin{align}
j_{\mathrm{reac}} &\sim \frac{Da_a D_{+} c_{\ce{H2O2}}}{c_{\ce{H2O2},\,0}}\\
j_{\mathrm{diff}} &\sim \frac{D_{\ce{H2O2}} \Delta c_{\ce{H2O2}}}{l} \label{eqn:diff_scaling}
\end{align}
where $l$ is a characteristic length scale over which the concentration of \ce{H2O2} changes by $\Delta c_{\ce{H2O2}}$, and $c_{\ce{H2O2},\,0}$ is its initial concentration. As a conservative choice for $l$, we choose the diffusion length $l \sim \sqrt[]{D_{\ce{H2O2}}t}$. This choice is conservative because additional mixing from the induced flow would only shorten the length scale. Substituting this value for $l$ into (\ref{eqn:diff_scaling}), equating the two fluxes, and solving for $t$ gives a scaling for the time associated with a given change in \ce{H2O2} concentration
\begin{align}
    \label{eqn:t_depletion}
    t_{\ce{H2O2}} \sim \left( \frac{\Delta c_{\ce{H2O2}}}{c_{\ce{H2O2}}} \right)^2 \frac{D_{\ce{H2O2}} c^2_{\ce{H2O2},\, 0}}{Da^2_a D^2_+}
\end{align}
In the cases investigated in the this study, over dynamically relevant times, which will be at worst $t\sim 1$ (the diffusion time across the domain), $\Delta c_{\ce{H2O2}}/c_{\ce{H2O2}}$ is small. Therefore, in effectively unbounded domains such as the current study, it is justifiable to ignore changes in \ce{H2O2} concentration over intermediate timescales which are long relative to system dynamics but short relative to timescales for depletion of \ce{H2O2} at the surface. For longer term behavior, one may adopt our solution as a quasi-steady solution in which the bulk concentration of \ce{H2O2} is adjusted gradually with time. In this case (\ref{eqn:t_depletion}) would imply that the length scale of depletion of \ce{H2O2} would be much longer than $\lambda$ while varying slowly in time. Another possibility is the presence of an effective reservoir \ce{H2O2} concentration, due to either bulk mixing or for nano-motors the motion of the swimmer itself, in which case a suitable \ce{H2O2} concentration may be assumed.

\subsection{Computational Framework}

We solve the governing equations using a custom written OpenMP parallel c++ code described thoroughly and compared against a commercial solver in \cite{Karatay2015}. The code uses second order staggered finite differences to discretize the governing equations where species concentrations, electric potential and pressure fields are stored at cell centers while velocities and fluxes are stored at cell faces. The linear Poisson and Poisson-like equations for pressure, electric potential, and momentum are solved using a pseudo-spectral method where fourier transforms are performed in the tangential $x$ direction and leaving a tridiagonal system to be solved at each wavenumber in the electrode-normal. Time integration is done with a semi-implicit method which iteratively solves the transport, momentum, and potential equations in a decoupled manner until the full implicit solution for the next timestep is reached. The species transport equations are solved implicitly during each iteration only in the stiff electrode-normal direction. Verification of the accuracy and convergence of the code was performed using the method of manufactured solutions~\cite{Salari2000}. This code and its variants have been successfully used in previous studies for prediction of chaotic electroconvection~\cite{Druzgalski2013, Druzgalski2016}, induced-charge electro-osmosis~\cite{Davidson2014}, and flow over patterned membranes~\cite{Davidson2016}. Mesh convergence testing was performed to ensure that a sufficiently fine mesh was used to resolve the results shown.

\section{Results and Analysis}

\begin{figure}[htb!]
	\centering
    \includegraphics[width=.8\columnwidth]{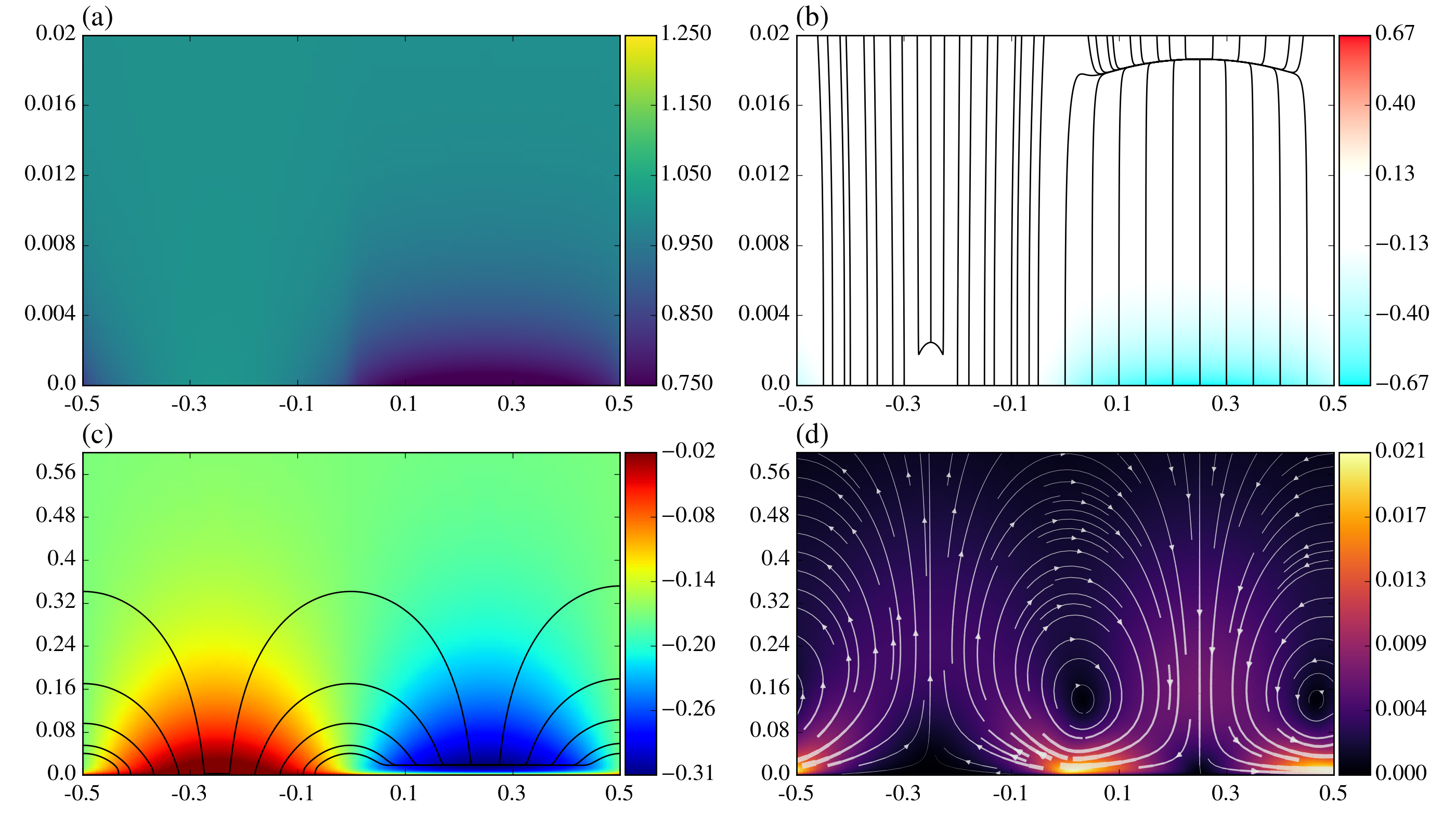}
    \caption{(Color online) Field quantities for $Da_a=1.48$, $Da_c=2.68$. (a) $H^+$ concentration field. (b) $\rho_e$ field with electric field lines. (c) $\phi$ field with electric field lines. (d) velocity magnitude field with streamlines.}
    \label{fig:qualitative}
\end{figure}

A number of dimensionless parameters are present in the above equations. We summarize these along with the values used in our simulations in table~\ref{table:params}. In this work, most of the parameters are held constant while two parameters, the surface Damk{\"o}hler numbers are varied. For reference to a physical system, the value $\epsilon=\num{4e-3}$ is representative of a .25mm pattern length $\lambda$ and pure water reservoir concentration of $c_0=\num{1e-7}$M. Qualitative steady-state fields of concentration, charge density, electric potential and electric field lines, and velocity magnitude with flow streamlines for nominal Damk{\"o}hler numbers of $Da_a=1.48$ and $Da_c=2.68$ are shown in figure~\ref{fig:qualitative}. These Damk{\"o}hler numbers come from rate constants $k_{o,\,anode}=\num{5.5e-9}\mathrm{m/s}$ and $k_{r,\,cathode}=1\,\mathrm{m^7s^{-1}mol^{-2}}$~\cite{MORAN2011} and peroxide concentration $c_{\ce{H2O2}}=\num{1e-3}$M. Evident from the concentration field in figure~\ref{fig:qualitative}a is that, outside of the double layer, the concentration of \ce{H+}, and also \ce{OH-} from electroneutrality, is equal to $c_0$. The combination of water equilibrium and electroneutrality prevent concentration gradients from forming in the bulk. This would change if other ionic species were present. The streamlines in figure~\ref{fig:qualitative}d show that a pair of counter-rotating vortices has been generated by the electric field visible in figure~\ref{fig:qualitative}c acting on the charged Debye layer shown in figure~\ref{fig:qualitative}b. Note the expanded $y$ scale in figure~\ref{fig:qualitative}a and b.

\begin{table}[!htb]
\centering
\begin{tabular}{c | c  | c | c}
  Parameter  & Description & Formula & Value(s) \\ \hline
  \rule{0pt}{2.5ex} 
  $\epsilon$ & dimensionless Debye length  & $\lambda_D/\lambda$     & $\num{4e-3}$ \\
  $Sc$       & Schmidt number              & $\mu/{\rho D_0}$        &  1000 \\
  $\kappa$   & Electrohydrodynamic coupling constant & $\varepsilon V_T^2/\mu D_0$ & .094 \\
             & domain aspect ratio         & $L/\lambda$             & 4 \\
  $D_+$      & \ce{H+} diffusivity         & $\widetilde{D}_+/D_0$   & 1.7658 \\
  $D_-$      & \ce{OH-} diffusivity        & $\widetilde{D}_-/D_0$   & 1 \\
  $Da_b$     &  bulk Damk{\"o}hler number  & $k_b\lambda^2c_0/D_0$ & $\num{9.38e4}$ \\
  $Da_a$     &  anode Damk{\"o}hler number & $k_{o,\,anode}\lambda c_{\ce{H2O2}}/D_+c_0$ & $\num{2.5e-3}$ \text{--} $\num{2.5e2}$\\
  $Da_c$     &  cathode Damk{\"o}hler number & $k_{r,\,cathode}\lambda c_0c_{\ce{H2O2}}/D_+$ & $\num{2.5e-3}$ \text{--} $\num{2.5e2}$
\end{tabular}
    \caption{Dimensionless governing parameters}
    \label{table:params}
\end{table}

\begin{figure}[htb!]
	\centering
    \includegraphics[width=1.\columnwidth]{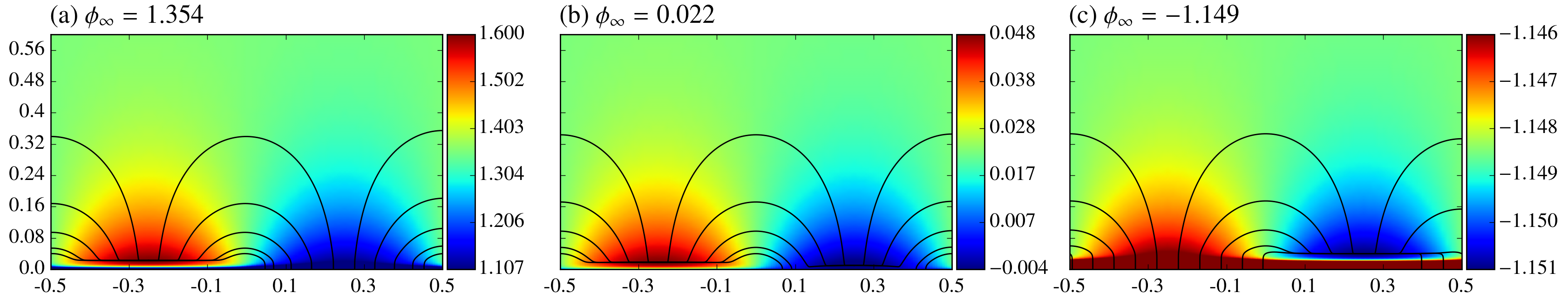}
    \caption{(Color online) Electrostatic potential fields with electric field lines plotted for several values of the Damk{\"o}hler numbers. (a) $Da_a=2.5$, $Da_c=.25$ (b) $Da_a=.25$, $Da_c=.25$ (b) $Da_a=.025$, $Da_c=.25$}
    \label{fig:potential_fields}
\end{figure}

One quantity of interest is the induced potential and particularly the induced potential difference between the surface and far field which we label $\phi_\infty$. Figure~\ref{fig:potential_fields} shows induced electric potential and associated electric field lines for different values of the anode Damk{\"o}hler number. The color scale in the plots have been set to emphasize variations about $\phi_\infty$ for each case. As the ratio of Damk{\"o}hler numbers changes, the sign of the induced net electric potential changes, and two patterns in the potential field are evident. First there is a 1D component associated with the net induced potential drop strongly visible within the double layer and varying only in the $y$ direction. There is also evident a component of the potential associated with the outer nearly semi-circular electric field lines which doesn't appear to be significantly affected by the changes in $Da_a$.

\subsection{An Approximate Model}

\begin{figure}[htb!]
	\centering
    \includegraphics[width=.7\columnwidth]{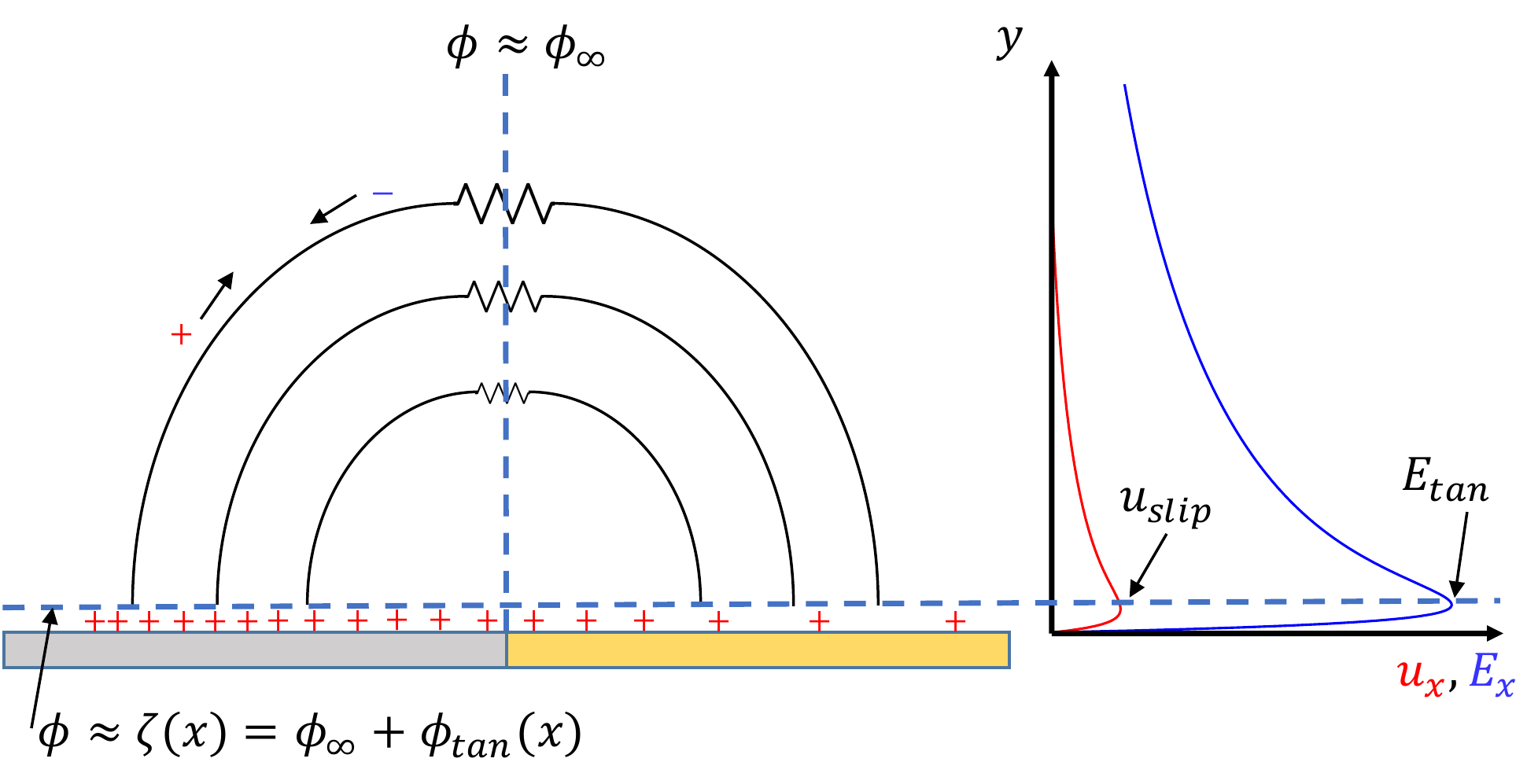}
    \caption{(Color online) Schematic showing conceptually our approximate model, and how we evaluate it against the numerical simulations. The zeta potential just outside the double layer here shown as positively charged. $\zeta(x)$ is calculated as a sum of two terms, the far-field potential $\phi_{\infty}$ and a component which varies along the surface $\phi_{\mathrm{tan}}(x)$. Current in the bulk travels along semi-circular lines which act as resistors. The model-predicted slip velocity $u_{\mathrm{slip}}$ and tangential electric field $E_{\mathrm{tan}}$ are compared against the maximum $x$ velocity and $x$ electric field along the line $x=0$.}
    \label{fig:model_schematic}
\end{figure}

In this section we develop an approximate model, shown schematically in figure~\ref{fig:model_schematic}, to understand the results of our numerical simulations. Our model utilizes balances of fluxes along with the approximation that current travels along semi-circular electric field lines outside of the double layer to calculate several important quantities which we compare against the results of our numerical simulations. One quantity is the induced potential difference between the surface and bulk $\phi_{\infty} = \phi(y \rightarrow \infty)$. A second is the reaction-induced zeta potential $\zeta(x) = \phi_{\infty} + \phi_{\mathrm{tan}}(x)$ where $\phi_{\mathrm{tan}}(x)$ describes the variation along the electrode surface just outside of the double layer of the electric potential and hence the tangential electric field responsible for inducing a slip velocity is $E_{\mathrm{tan}} = \frac{d\zeta}{dx}=\frac{d\phi_{\mathrm{tan}}}{dx}$. These allow us to use the Smoluchowski equation to calculate a third quantity, the slip velocity $u_{\mathrm{slip}}$.

We start with a scaling which helps explain the behavior of $\phi_{\infty}$. At steady state, the total reactive currents at the anode and cathode must balance. At the anode, the reaction rate in our model is determined solely by the constant concentration of \ce{H2O2}. This means that the system must adjust the reaction rate at the cathode to conserve current. The only way that this adjustment can occur is by changing the surface \ce{H+} concentration to an appropriate average value via an induced potential $\phi_{\infty}$. We label this estimate of the induced potential $\phi_{\infty, \, 0}$ because we will later calculate $\phi_{\infty}$ in a more complete way accounting for transport between anode and cathode, and $\phi_{\infty, \, 0}$ will show up as the first term in that equation. When $c_+$ obeys a Boltzmann distribution, valid for small currents, and is equal to the reservoir concentration outside of the double layer then
\begin{align}
    \label{eqn:j_pt}
    j^+_{\ce{Pt}} &= D_+ Da_a \\
    j^+_{\ce{Au}} &= \left(e^{\phi_{\infty,\,0}}\right)^2D_+Da_c
\end{align}
where $j^{+}_{i}$ is the magnitude of the reactive surface flux. Setting these equal and solving for  $\phi_{\infty,\,0}$ gives
\begin{align}
    \phi_{\infty,\,0} = 1/2\ln\left(Da_a/Da_c\right)
\end{align}
Physically, the dependence of $\phi_{\infty,\,0}$ on the ratio of Damk{\"o}hler numbers can be interpreted as the potential adjusting to either increase or decrease the surface concentration of \ce{H+} in such a way as to balance the reaction rates at the electrodes. This leads to a dependence on the ratio of reaction rate constants and the bulk \ce{H+} concentration, but not on the \ce{H2O2} concentration.

In order to understand the tangential component of the electric field we look at the transport of ions from anode to cathode. In the absence of significant bulk charge, as is evident in figure~\ref{fig:qualitative}(b), current is carried predominantly in an Ohmic manner. Equating scaling of electromigration current and the cathodic current gives

\begin{align}
    E_{\mathrm{tan}} \sim Da_a
\end{align}

In addition to governing transport of cations from anode to cathode, this component of the electric field will also affect the average zeta potential of the cathode. In other words, even if $Da_a=Da_c$ and thus $\phi_{\infty,\,0}=0$, the fact that the potential at the anode outside of the double layer must be higher than the potential at the cathode to drive transport from one to the other means that $\phi_{\infty}$ must be induced to equalize the average reaction rates. This contribution to the net potential will scale as the electric field multiplied by the length scale over which it acts. We estimate the coefficients associated with this scaling and account for the quadratic dependence of the reaction rate on the \ce{H+} concentration at the cathode using a model based on Ramos' work~\cite{Ramos2003} on Alternating Current Electro-Osmosis. We assume that inside of the double layers all transport is in the electrode-normal direction while outside of them current travels along semi-circular electric field lines. This gives a bulk current of
\begin{align}
    \label{eqn:i_bulk}
    i=\left(D_{+} + D_{-}\right)\frac{\phi_{\mathrm{tan}}(-x) - \phi_{\mathrm{tan}}(x)}{\pi x}
\end{align}
where $i$ is the current between anode and cathode following the semi-circular electric field line starting at $-x$, ending at $x$, and just outside the electric double layer. (\ref{eqn:i_bulk}) is valid in the range $0<x<1/4$ where current lines travel from the anode to the right to the cathode without passing through the periodic boundary. We restrict our analysis to this region for simplicity and without loss of generality because the problem is symmetric about the center of each electrode. We will use (\ref{eqn:j_pt}) along with (\ref{eqn:i_bulk}) to predict an $i$ which is constant along the surface. While this is not generally true at the cathode where $i$ depends on $c_+$ and therefore is a function $x$, this model provides a useful approximation
\begin{align}
    \label{eqn:dphi_tan}
	\phi_{\mathrm{tan}}\left(x\right) = \frac{-Da_a \pi x}{2\left(1 + D_-/D_+\right)}
\end{align}

\begin{figure}[htb!]
	\centering
    \includegraphics[width=.8\columnwidth]{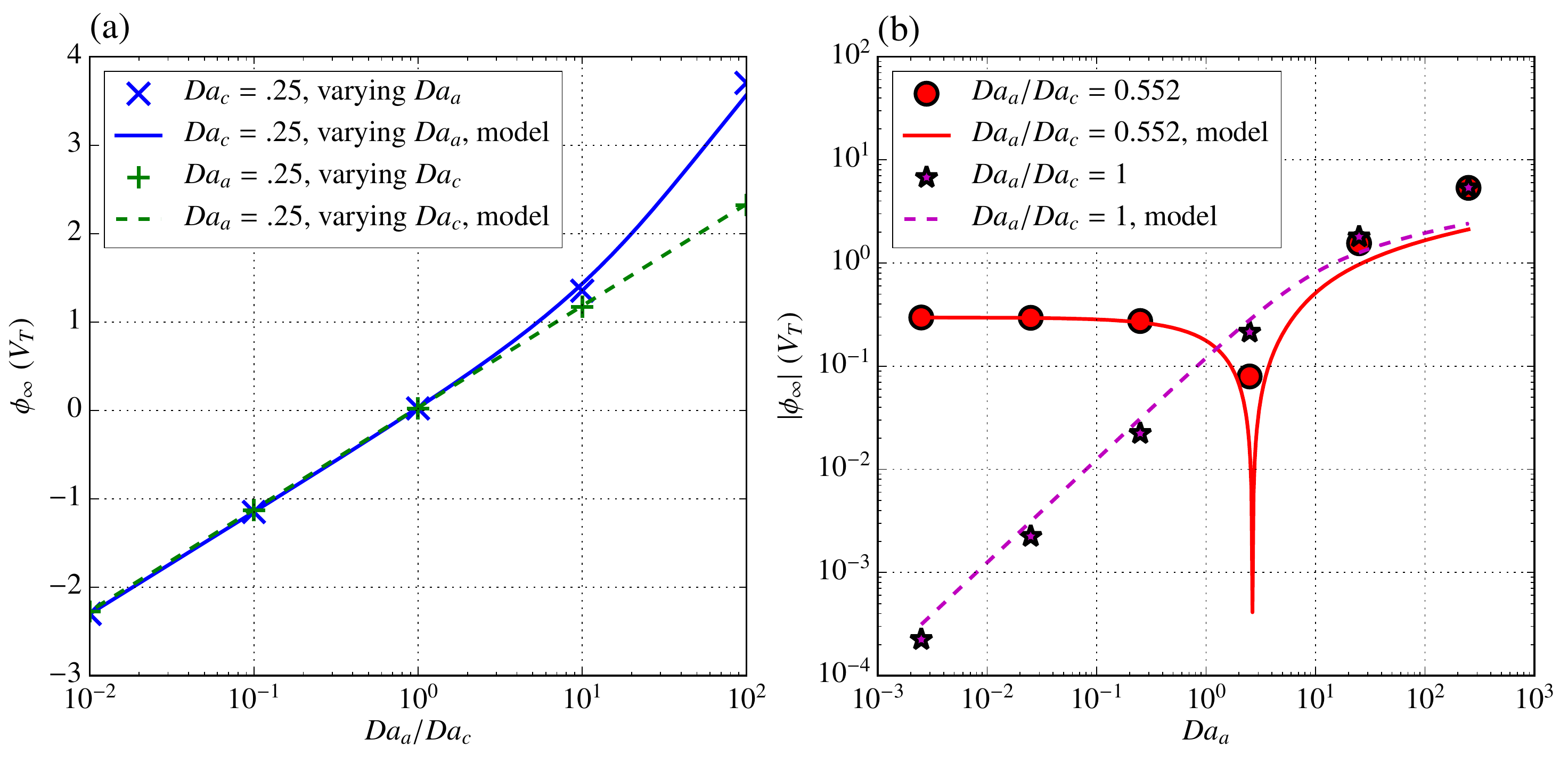}
    \caption{(Color online) Scaling of $\phi_{\infty}$ and comparison of simulation with model (\ref{eqn:phi_infty}). (a) Scaling with Damk{\"o}hler ratio showing the predicted logarithmic scaling when $Da_a$ is small. (b) Scaling with $Da_a$ for two fixed values of $Da_a/Da_c$.}
    \label{fig:net_potential}
\end{figure}

Now that we have an estimate for the tangential variation of the electric potential, we can calculate an estimate of $\phi_{\infty}$ which accounts for it. Previously, to find $\phi_{\infty,\,0}$, we assumed that the potential drop across the double layer, and hence the surface concentration and reaction flux were independent of $x$. We now relax this assumption and include the combined potential difference across the double layer $\zeta = \phi_{\infty} + \phi_{\mathrm{tan}}$. Setting the total flux at the cathode equal to that at the anode by integrating over the surface gives 
\begin{align}
    \int_{0}^{1/4} \left(e^{\zeta} \right)^2 D_+ Da_c dx = \frac{\lambda}{4} D_+Da_a
\end{align}
where we have restricted our integral to be over the half of the cathode where (\ref{eqn:dphi_tan}) gives a valid estimate for the contribution of transport to the potential difference across the double layer. By symmetric, the current through the other half of the cathode will come from the other half of the anode. Solving for $\phi_{\infty}$ gives 
\begin{align}
    \label{eqn:phi_infty}
	\phi_{\infty} = \phi_{\infty,\, 0} + \frac{1}{2}\ln\left(
		\frac{\beta Da_a}{1 - \exp{\left(-\beta Da_a\right)}}
        \right)
\end{align}
where $\beta = \pi / 4 \left( 1 + D_-/D_+ \right)$. In order to evaluate these models and scalings, we performed simulations over a broad range of anode and cathode Damk{\"o}hler numbers. Figure~\ref{fig:net_potential} shows a comparison between $\phi_{\infty}$ from those simulations and the results of our model.

In figure~\ref{fig:net_potential}a cases where one of the Damk{\"o}hler numbers is fixed and the other varied are shown, and very good agreement is observed over the entire range studied. Experimentally varying this ratio would be difficult because it depends on the ratio of reaction rate coefficients, however some variation may be possible by choosing different electrode materials. Figure~\ref{fig:net_potential}b shows cases where the ratio of the Damk{\"o}hler numbers is fixed and their magnitude is varied together representative of varying the concentration of \ce{H2O2}. Over most of the parameter space good agreement is observed except when $Da_a$ becomes very large $O(100)$ and the model assumptions used to determine the component of $\phi_{\infty}$ caused by tangential transport break down. Additionally, at $Da_a=2.5$ and a ratio of $Da_a/Da_c=.552$, some error is apparent because $\phi_{\infty}$ changes sign near this point making the result on a log-scale highly sensitive to the exact Damk{\"o}hler number at which this sign change occurs. Both of these plots demonstrate that despite the many assumptions in the model, it is able to accurately capture not merely the scaling, but the induced net potential in a broad parameter space.

\begin{figure}[htb!]
	\centering
    \includegraphics[width=.8\columnwidth]{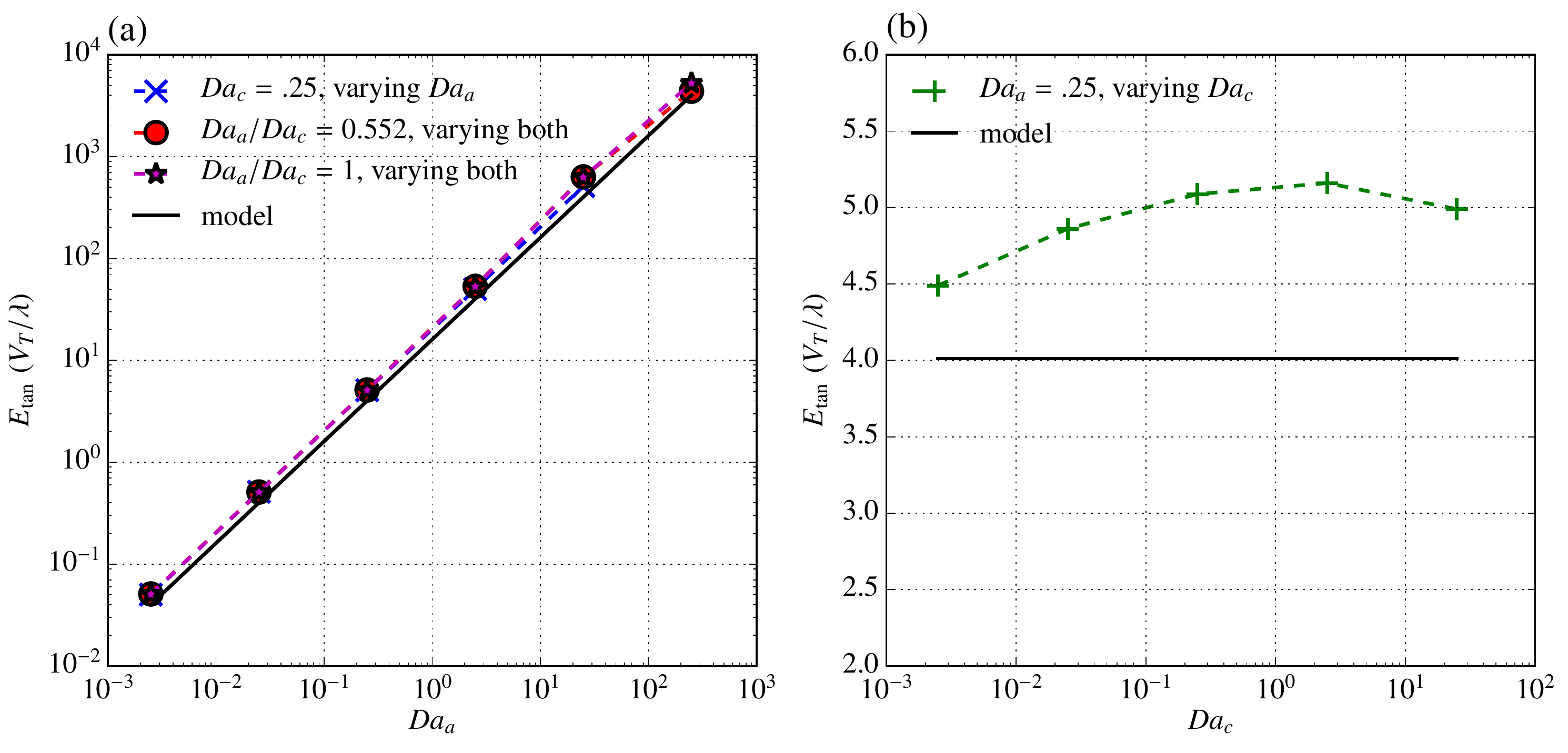}
    \caption{(Color online) Scaling of tangential electric field with Damk{\"o}hler numbers. (a) shows that as predicted $E_{\mathrm{tan}}$ scales linearly with $Da_a$. (b) shows the independence of $E_{\mathrm{tan}}$ from $Da_c$.}
    \label{fig:E_tan}
\end{figure}

To evaluate our predicted linear scaling of the tangential electric field with $Da_a$, figure~\ref{fig:E_tan} shows a comparison of our model
\begin{align}
    \label{eqn:E_tan}
    E_{\mathrm{tan}}=2\beta Da_{a}
\end{align}
with $E_{\mathrm{tan}}$ from the simulations defined as $\max\left({E_x(x=0,y)}\right)$. We again see that the scaling is quite accurate over the entire range simulated with the predicted linear behavior when $Da_a$ is varied and almost no change when $Da_c$ is varied while $Da_a$ is held constant. Additionally, our simple model is able to estimate the magnitude of the tangential electric field to within an $O(1)$ constant over the entire range.

Our model for the electric field (\ref{eqn:E_tan}) predicts that it should depend on the diffusivities of both species. In contrast, previous studies~\cite{Esplandiu2016} found that it should depend only on the ionic strength and \ce{H+} diffusivity. They argue that the conductivity shouldn't be important because only \ce{H+} ions carry net current. However, when bulk water equilibrium reactions are accounted for, current in the bulk can be carried by \ce{OH-} ions which are generated by water splitting near the cathode and recombine with \ce{H+} ions near the anode leading to a dependence of the electric field on their diffusivity.

\begin{figure}[htb!]
	\centering
    \includegraphics[width=.8\columnwidth]{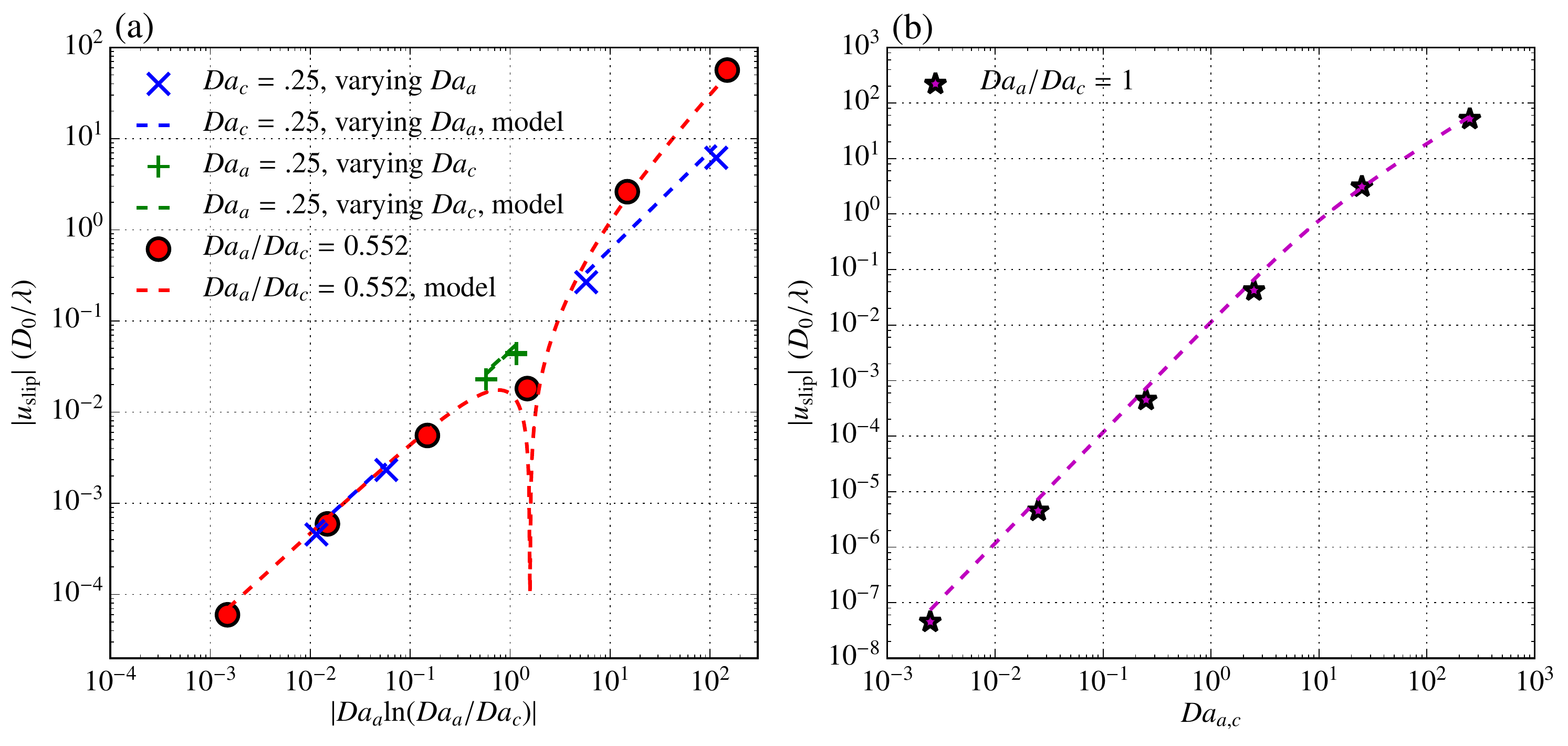}
    \caption{(Color online) Comparison of maximum tangential velocity predicted by our model (dashed lines) with simulations (symbols). (a) comparison for $Da_a/Da_c \neq 1$. (b) comparison for $Da_a/Da_c = 1$.}
    \label{fig:vel_scaling}
\end{figure}

The velocity of the flow induced by the micropump provides a useful measure of the ability of the reaction-induced convection to provide additional transport on top of diffusion. From a balance of the electric body force and diffusive terms in the NS equations, the velocity should scale as the product of $\zeta$ and $E_{\mathrm{tan}}$. For our model, we use the well known Smoluchowski formula for electroosmotic flow past a flat charged surface, $u_{slip} = \varepsilon E \zeta/\mu$ which in our dimensionless formulation gives
\begin{align}
    u_{slip}(x) =& \kappa E_{\mathrm{tan}}\zeta(x)\\
                =& \kappa \beta Da_{a}
             	   \left(\ln{\frac{Da_a}{Da_c}}
                   + \ln{\frac{\beta Da_a}{1 - \exp\left({-\beta Da_a}\right)}}
                   - 2 \beta Da_a x\right)
\end{align}
We evaluate this predicted slip velocity by comparing the predicted velocity at $x=0$ with the slip velocity from our simulations $\max\left(u_x(x=0,y)\right)$ in figure~\ref{fig:vel_scaling}. When $\phi_{\infty, \, 0}$ is dominant, our model predicts that the velocity should be proportional to the quantity $Da_{a}\ln\left(Da_a/Da_c\right)$. From figure~\ref{fig:vel_scaling}a we see that the velocity does indeed scale with this quantity and that our model predicts the maximum simulated velocity to within an $O(1)$ in all cases. In figure~\ref{fig:vel_scaling}b a comparison is shown for the case where $Da_a/Da_c=1$ and hence $\phi_{\infty,\,0}=0$. Again for this case agreement to within an $O(1)$ constant is seen over many orders of magnitude change in the Damk{\"o}hler number.

\section{Discussion and Conclusions}
In this work we have modeled the reaction-induced electroconvection generated by the electrochemical decomposition of hydrogen peroxide on platinum and gold surfaces. Based on observations from our direct numerical simulations, we have developed a simple model which is able to accurately captured both the induced electric potentials and flow velocity magnitudes over a broad range of Damk{\"o}hler numbers. This model can be used to aid in experimental design of systems using this reaction induced flow to provide additional near-surface mixing. It can also be easily modified to provide estimations for systems which involve other electrochemical reactions.


\section{Acknowledgements}
Scott M. Davidson was supported by a Robert and Katherine Eustis Stanford Graduate Fellowship and by the National Science Foundation Graduate Research Fellowship under Grant No. DGE-114747. Ali Mani and Scott M. Davidson were supported by the National Science Foundation under Award \#1553275. Rob G.H. Lammertink would like to acknowledge the Netherlands Organisation for Scientific Research (NWO, Vici project “stirring the boundary layer”).

\bibliographystyle{unsrt}

\end{document}